\documentclass[12pt]{article}

\usepackage{graphicx}
\usepackage{scicite}
\usepackage{times}
\newenvironment{sciabstract}{%
\begin{quote} \bf}
{\end{quote}}

\newcounter{lastnote}
\newenvironment{scilastnote}{%
\setcounter{lastnote}{\value{enumiv}}%
\addtocounter{lastnote}{+1}%
\begin{list}%
{\arabic{lastnote}.} {\setlength{\leftmargin}{.22in}}
{\setlength{\labelsep}{.5em}}} {\end{list}}

\title{Universal Quantum Viscosity in a Unitary Fermi Gas}

\author{C. Cao, E. Elliott, J. Joseph, H. Wu, J.~Petricka$^1$, T. {Sch\"{a}fer}$^2$, and J.~E.~Thomas$^{\ast}$\\
\\
\normalsize{Physics Department, Duke University, Durham, North
Carolina 27708}\\
\normalsize{$^1$Physics Department, Gustavus Adolphus College,
Saint Peter, Minnesota 56082}\\
\normalsize{$^2$Physics Department, North Carolina State University, Raleigh, North Carolina 27695}\\
\\
\normalsize{$^\ast$To whom correspondence should be addressed;
E-mail:  jet@phy.duke.edu.}\\
\\
\normalsize{  }\\
\normalsize{}}
\date{}

\begin{document}


\baselineskip24pt


\maketitle


\begin{sciabstract}
A Fermi gas of atoms with resonant interactions is predicted to obey
universal hydrodynamics, where the shear viscosity and
other transport coefficients are universal functions of the
density and temperature.  At low temperatures, the viscosity has a universal quantum scale $\hbar n$ where $n$ is the density, while at high temperatures the natural scale is  $p_T^3/\hbar^2$ where $p_T$ is the thermal momentum. We employ breathing mode damping to measure the shear viscosity at low temperature. At high temperature $T$, we employ anisotropic expansion of the cloud to find the viscosity, which exhibits precise $T^{3/2}$ scaling. In both experiments, universal  hydrodynamic equations including friction and heating are used to extract the viscosity. We  estimate the ratio of the shear viscosity to the entropy density and compare to that of a perfect fluid.
\end{sciabstract}

Ultracold strongly interacting Fermi gases are of broad interest,
as they provide a tunable tabletop paradigm  for
strongly interacting systems, ranging from high temperature
superconductors to nuclear matter. First observed in 2002, quantum degenerate, strongly
interacting Fermi gases are being widely
studied~\cite{OHaraScience,RMP2008,ZwergerReview,ZwierleinFermiReview}. To obtain strong
interactions (characterized by a divergent s-wave scattering length), a bias magnetic field is used to tune
the gas to a broad collisional (Feshbach) resonance, where the
range of the collision potential is small compared to the
interparticle spacing.
In this so-called unitary regime,  the properties of the gas are
universal functions of  the density $n$ and temperature $T$. The universal behavior of the equilibrium thermodynamic properties has been studied in
detail~\cite{JointScience,LuoEntropy,JinPotential,DrummondUniversal,ThermoLuo,ThermoUeda,ThermoSalomon}, whereas the measurement of universal transport coefficients presents new
challenges.

The measurement of the viscosity is of particular interest in the
context of a recent conjecture, derived using string theory methods,
which defines a perfect normal fluid~\cite{Kovtun}. An example of a nearly perfect fluid is the quark-gluon plasma produced in gold ion collisions, which exhibits almost perfect frictionless flow and is thought to be a good approximation to the state of matter that existed microseconds after the Big Bang~\cite{McClerranRHIC}.  The conjecture states that the ratio of the shear viscosity $\eta$ to the entropy density $s$
has a universal minimum, $\eta/s\geq \hbar/(4\pi k_B)$.  This ratio is experimentally accessible in a trapped unitary Fermi gas, where the entropy has been measured both globally~\cite{LuoEntropy,ThermoLuo} and locally~\cite{ThermoUeda,ThermoSalomon} and the viscosity can be determined from hydrodynamic experiments~\cite{ShuryakQuantumViscosity,BruunViscousNormalDamping,SchaferRatio,TurlapovPerfect}, so that the predicted minimum ratio can be directly compared to that from Fermi gas experiments~\cite{SchaferRatio,TurlapovPerfect}.

In a Fermi gas, the $\eta/s$ ratio for the normal fluid is expected to reach a minimum just above the superfluid transition temperature~\cite{SchaferRatio}.
This can be understood using dimensional analysis.
Shear viscosity has units
of momentum/area. For a unitary gas, the natural momentum is the relative momentum $\hbar k$ of a colliding pair of particles, whereas the natural area is the
resonant s-wave collision cross section~\cite{p-wave}, $4\pi/k^2$. Thus, $\eta \propto \hbar k^3$.
At temperatures well below the Fermi temperature at which
degeneracy occurs, the Fermi momentum sets the scale so
$k\simeq 1/L$, where $L$ is the interparticle spacing. Then
$\eta\propto\hbar /L^3$ and $\eta\propto\hbar n$.
For a normal fluid above the critical temperature, the scale of entropy
density $s\simeq n\,k_B$, so  $\eta/s\simeq\hbar /k_B$.
For much higher temperatures above the Fermi temperature, one expects
that $\hbar k$ is comparable to the thermal momentum
$p_T=\sqrt{2mk_BT}$, giving the scale $\eta\propto p_T^3/\hbar^2\propto T^{3/2}/\hbar^2$.

To properly  measure the shear viscosity with high precision over a wide temperature range, we use universal hydrodynamic equations, which contain both the friction force and the heating rate, to extract the viscosity from two experiments, one for each of two temperature ranges. For measurement at high temperatures, we observe the expansion dynamics of a unitary Fermi gas after release from a deep optical trap and demonstrate the predicted universal $T^{3/2}$ temperature scaling.   For measurement at low temperatures, we employ the damping rate of the radial breathing mode,  using the raw cloud profiles from our previous work~\cite{KinastDampTemp}. The smooth joining(discontinuity) of the data from the two measurement methods when heating is included(excluded)~\cite{SupportOnline},  demonstrates the importance of including the heating as well as the friction force in the universal hydrodynamic analysis.

The experiments employ a 50-50 mixture of the two lowest hyperfine states of $^6$Li, which is magnetically tuned to a broad Feshbach resonance and cooled by evaporation in the optical trap. The initial energy per particle $E$ is measured from the trapped cloud profile~\cite{SupportOnline}.

In the  high temperature regime,  the total
energy of the gas $E$ is larger than $2\,E_F$, well above the
critical energy $E_c<0.8\,E_F$ for the superfluid
transition~\cite{ThermoLuo,ThermoUeda,ThermoSalomon}. In this
case, the density profile is well fit by a Gaussian,
$n(x,y,z,t)=n_0(t)\,\exp(-x^2/\sigma_x^2-
y^2/\sigma_y^2-z^2/\sigma_z^2)$, where $\sigma_i(t)$ is a time
dependent width  and
$n_0(t)=N/(\pi^{3/2}\sigma_x\sigma_y\sigma_z)$ is the central
density and $N$ is the total number of atoms.

\begin{figure}
\begin{center}\
(A)
\includegraphics[width=150mm]{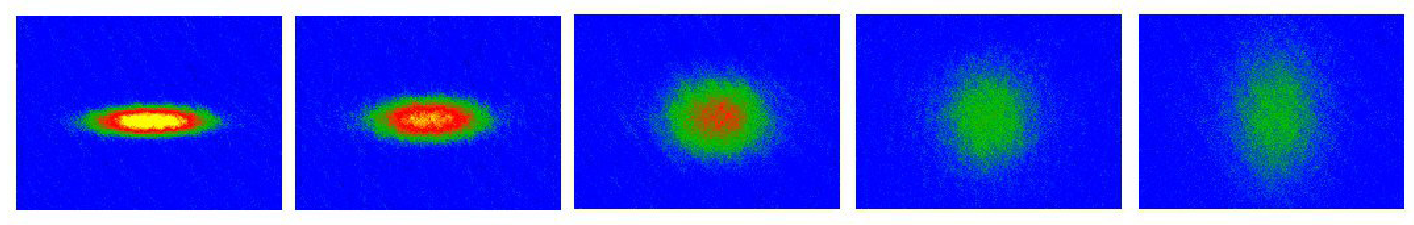}

(B)
\includegraphics[width=120mm]{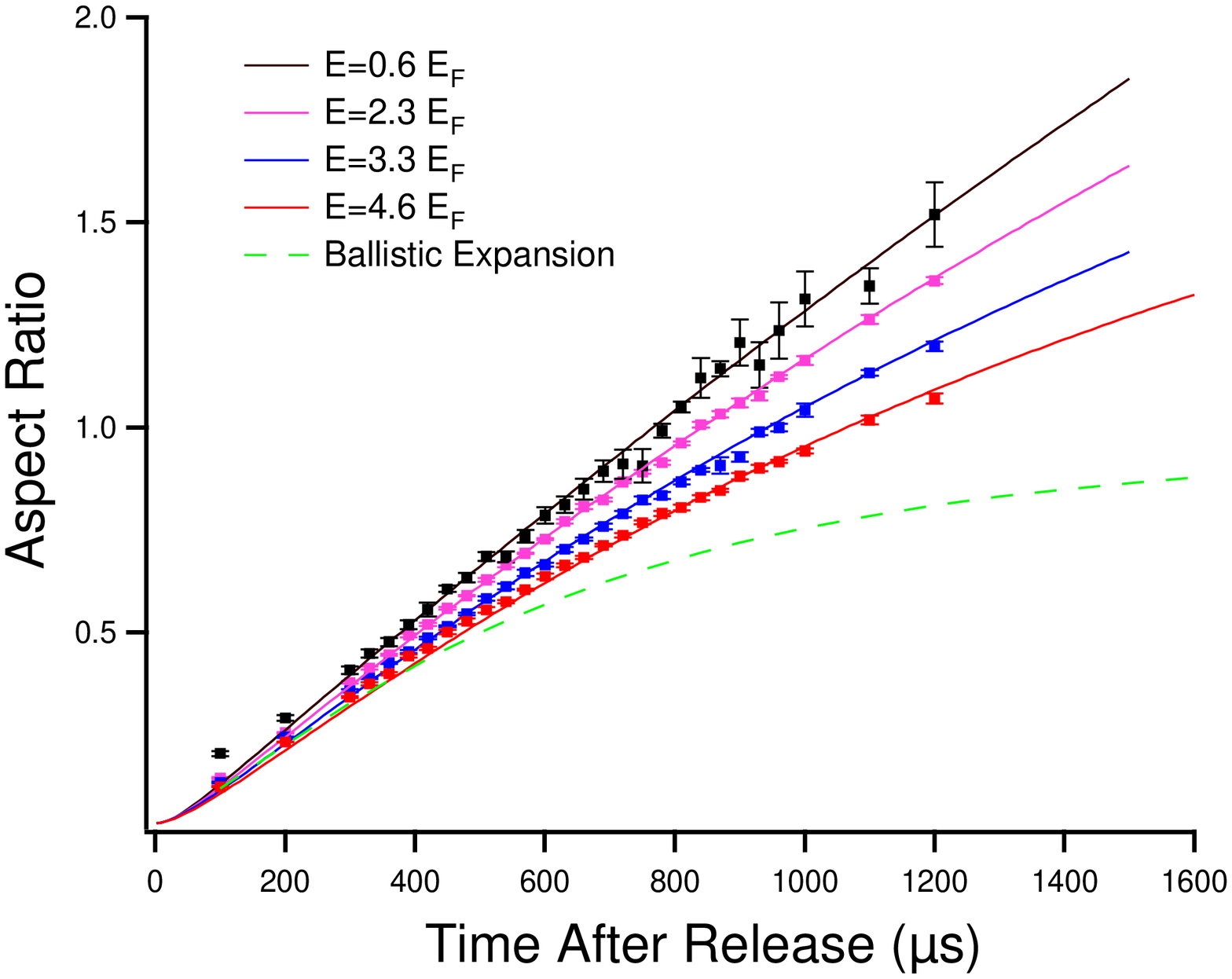}
\end{center}
\caption{Anisotropic expansion. (A) Cloud absorption images for 0.2, 0  .3, 0.6, 0.9, 1.2 ms expansion time, $E=2.3\,E_F$; (B) Aspect ratio versus time. The expansion rate decreases at higher energy as the viscosity increases. Solid curves: Hydrodynamic theory with the viscosity as the fit parameter.  Error bars denote statistical fluctuations in the aspect ratio.\label{fig:aspectratio}}
\end{figure}

The  aspect ratio $\sigma_x(t)/\sigma_z(t)$ is measured as a
function of time after release to characterize the hydrodynamics,
 for different energies $E$ between $2.3\,E_F$
and $4.6\,E_F$, Fig.~\ref{fig:aspectratio}. We also take expansion data at one low energy point $E=0.6\, E_F$, where the viscosity is small compared to that obtained at higher temperatures and the density profile is approximately a zero temperature Thomas-Fermi distribution. The black curve shows the fit for zero viscosity and no free parameters. To obtain a high signal to background ratio, we measure the aspect ratio only up to 1.4. For comparison, the green dashed curve shows the prediction for a ballistic gas.

We determine the shear viscosity $\eta$ by using a hydrodynamic
description of the velocity field $\mathbf{v}(\mathbf{x},t)$ in
terms of the scalar pressure and the shear viscosity pressure
tensor,
\begin{equation}
m\left(\partial_t
+\mathbf{v}\cdot\nabla\right)v_i=f_i + \sum_j\frac{\partial_j (\eta\,\sigma_{ij})}{n},
\label{eq:force}
\end{equation}
where $\mathbf{f}=-\nabla P/n$ is the  force per particle arising
from the scalar pressure $P$ and $m$ is the atom mass. For a unitary gas, the bulk
viscosity is predicted to
vanish in the normal fluid~\cite{SonBulkViscosity,EscobedoBulkViscosity}, so we do not
include it in the analysis for the expansion. The second term on the right describes the friction forces
arising from the shear viscosity, where $\sigma_{ij}=\partial
v_i/\partial x_j+\partial v_j/\partial
x_i-2\delta_{ij}\nabla\cdot\mathbf{v}/3$ is symmetric and traceless.

For a unitary gas, the  evolution equation for the pressure takes
a simple form, since $P=2{\cal E}/3$~\cite{HoUniversalThermo,
ThomasUniversal}, where ${\cal E}$ is the local energy density
(sum of the kinetic and interaction energy). Then, energy conservation and Eq.~\ref{eq:force} implies $(\partial_t
+\mathbf{v}\cdot\nabla + 5\nabla\cdot\mathbf{v}/3)P=2\dot{q}/3$.
Here, the heating rate per unit volume $\dot{q}=\eta\,\sigma_{ij}^2/2$ arises from friction due to the relative motion of neighboring volume elements.  To express this in terms of the force per particle, $f_i$, we differentiate this equation for $P$ with respect to $x_i$, and use the continuity equation for the density to obtain
\begin{equation}
 \left(\partial_t
+\mathbf{v}\cdot\nabla
+\frac{2}{3}\nabla\cdot\mathbf{v}\right)f_i + \sum_j(\partial_i v_j)f_j -\frac{5}{3}\left(\partial_i\nabla\cdot\mathbf{v}\right)\frac{P}{n}=-\frac{2}{3}\frac{\partial_i\dot{q}}{n}.
\label{eq:forceperparticle}
\end{equation}
 Force balance in the trapping potential $U_{trap}(\mathbf{x})$, just before release of the cloud, determines the initial condition $f_i(0)=\partial_i U_{trap}(\mathbf{x})$.

These hydrodynamic equations include both the force and the heating arising from viscosity.  The solution is greatly simplified when the cloud is released from a deep, nearly harmonic trapping potential $U_{trap}$, as $f_i(0)$ is then linear in the spatial coordinate. If we neglect viscosity, the force per particle and hence the
velocity field remain linear functions of the spatial coordinates as the cloud expands.
Thus $\partial_i(\nabla\cdot\mathbf{v})=0$ and the pressure $P$
does not appear in Eq.~\ref{eq:forceperparticle}. By numerical integration~\cite{ThomasNUM}, we find that non-linearities in the velocity field are very small even if the viscosity is not zero, because dissipative forces tend to restore a linear flow profile. Hence,  the evolution equations~\ref{eq:force} and~\ref{eq:forceperparticle}, are only weakly dependent on the precise initial spatial profile of $P$ and  independent of the detailed thermodynamic properties.

We therefore assume that the velocity field is exactly linear in the spatial
coordinates. We take $f_i=a_i(t)x_i$ and
 $\sigma_i(t)=b_i(t)\sigma_i(0)$, i.e., the density changes by a
scale transformation~\cite{Menotti}, where current conservation then
requires $v_i=x_i\,\dot{b}_i(t)/b_i(t)$.

In general, the viscosity takes the universal form $\eta = \alpha(\theta)\,\hbar n$, where $\theta$ is the local reduced temperature and $\eta\rightarrow 0$ in the low density region of the cloud~\cite{BruunViscous,SupportOnline}. Using the measured trap frequencies, and eqs.~\ref{eq:force} and~\ref{eq:forceperparticle}, the aspect ratio data are  fit to determine  the trap averaged viscosity parameter, $\bar{\alpha}=(1/N\hbar)\int d^3\mathbf{x}\,\eta(\mathbf{x},t)$, which arises naturally, independent of the spatial profile of $\eta$. Since $\theta$  has a zero convective derivative everywhere (in the zeroth order adiabatic approximation) and the number of atoms in a volume element is conserved along a stream tube, $\bar{\alpha}$ is a constant that can be compared to predictions for the trapped cloud before release.

 As shown in Fig.~\ref{fig:aspectratio}, the expansion data are very well fit over the range of energies studied, using $\bar{\alpha}$ as the only free parameter.  We find that the friction force produces a curvature that matches the aspect ratio versus time data, while the indirect effect of heating is significant in increasing the outward force, which increases the fitted $\bar{\alpha}$ by a factor of $\simeq 2$, compared to that obtained when heating is omitted~\cite{SupportOnline}.

 For measurements at low temperatures, where the viscosity is small, we determine $\bar{\alpha}$ from the damping rate of the radial breathing mode~\cite{KinastDampTemp}.  For the breathing mode, the cloud radii change by a scale transformation of the form $b_i=1+\epsilon_i$, with $\epsilon_i<<1$, and the heating rate in eq.~\ref{eq:forceperparticle} is  $\propto\dot{\epsilon}_i^2$, which is negligible. Hence, the force per particle evolves adiabatically. Adding the trapping force to eq.~\ref{eq:force}, one obtains the damping rate $1/\tau=\hbar\bar{\alpha}/(3m\langle x^2\rangle)$~\cite{SupportOnline,DampNote}.

The fitted viscosity coefficients $\bar{\alpha}$ for the entire energy range are shown
in Fig. 2, which can be used to test predictions~\cite{LevinViscosity,TaylorViscosity,ZwergerViscosity}.
Despite the large values of $\bar{\alpha}$ at the higher energies, the viscosity causes only a moderate perturbation to the adiabatic expansion, as shown by the expansion data and the fits in Fig.~\ref{fig:aspectratio}.
The breathing mode data and expansion data smoothly join, provided that the heating rate is included in the analysis. In contrast, omitting the heating rate  produces a discontinuity between the high and low temperature viscosity data~\cite{SupportOnline}. The agreement between these very different measurements when heating is included shows that hydrodynamics in the universal regime is well described  by eqs.~\ref{eq:force} and~\ref{eq:forceperparticle}.

To test the prediction of the $T^{3/2}$ temperature scaling in the
high temperature regime, we assume that $\eta$ relaxes to the
equilibrium value in the center of the trap, but vanishes in the low
density region so that $\bar\alpha$ is well defined. This behavior
is predicted by kinetic theory~\cite{BruunViscous}. We  expect that $\bar{\alpha}\simeq
\alpha_0$ where $\eta_0=\alpha_0 \hbar n_0$ is the viscosity at the trap center before release.  At high temperatures~\cite{BruunViscousNormalDamping},
\begin{equation}
\alpha_0 = \alpha_{3/2}\, \theta_0^{3/2},
\label{eq:scaling}
\end{equation}
where $\alpha_{3/2}$ is a universal coefficient.  As  $\theta$ has a zero convective
derivative everywhere (in the zeroth order adiabatic
approximation), $\theta_0$ at the trap center has a zero time
derivative and $\alpha_0$ is therefore constant as is $\bar{\alpha}$.

\begin{figure}
\begin{center}\
\includegraphics[width=120mm]{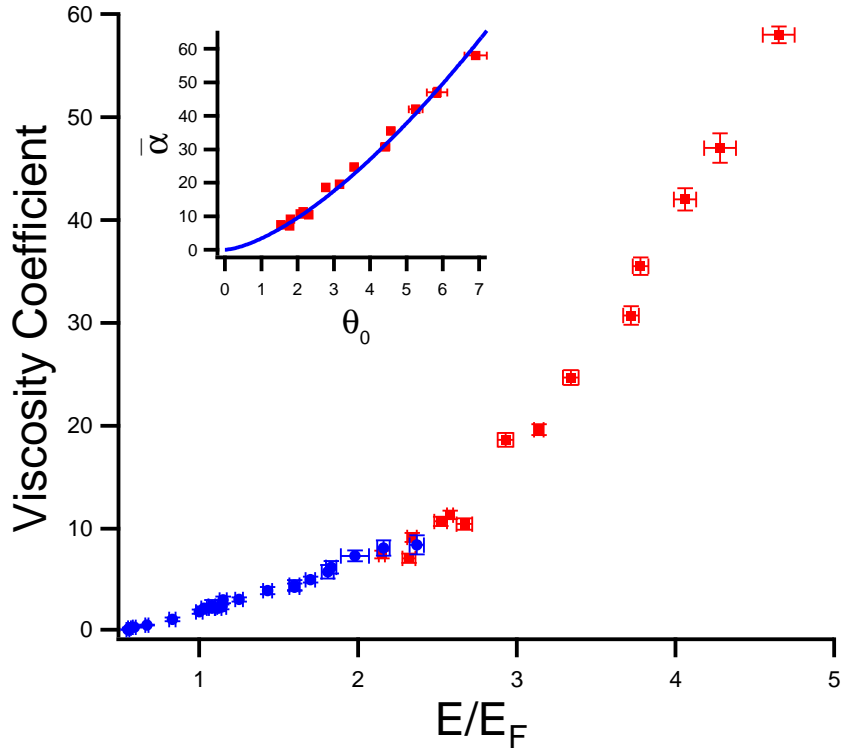}
\end{center}
\caption{Trap averaged viscosity coefficient $\bar{\alpha}=\int d^3\mathbf{x}\,\eta/(\hbar N)$ versus initial energy per atom. Blue circles: Breathing mode measurements; Red squares: Anisotropic expansion measurements. Bars denote statistical error arising from the uncertainty in $E$ and the cloud dimensions. Inset: $\bar{\alpha}$  versus reduced temperature $\theta_0$ at the trap center prior to release of the cloud.  The blue curve shows the  fit $\alpha_0=\alpha_{3/2}\,\theta_0^{3/2}$, demonstrating the predicted universal high temperature scaling. Bars denote statistical error arising from the uncertainty in $\theta_0$ and $\bar{\alpha}$. A 3\% systematic uncertainty in $E_F$ and 7\% in $\theta_0$ arises from the systematic uncertainty in the absolute atom number~\cite{SupportOnline}. \label{fig:viscosity}}
\end{figure}

The inset in Fig.~\ref{fig:viscosity} shows the high temperature (expansion) data
for $\bar\alpha$ versus the initial reduced temperature at the trap
center, $\theta_0$. Here, $\theta_0 = T_0/T_F(n_0) = (T_0/T_{FI})
(n_I/n_0)^{2/3}$. The local Fermi temperature $T_F(n_0) = \hbar^2
(3\pi^2n_0)^{2/3}/(2mk_B)$ and $T_{FI} = E_F /k_B = T_F (n_I)$ is
the ideal gas Fermi temperature at the trap center.  $n_I$ is
the ideal gas central density for a zero temperature Thomas-Fermi
distribution. We use $(n_I/n_0)^{2/3} = 4(\sigma^2_z/\sigma^2_{Fz})/
\pi^{1/3}$ and obtain the initial $T_0/T_{FI}$ from the cloud profile~\cite{SupportOnline}.

 The excellent fit of Eq.~\ref{eq:scaling} to the data, inset Fig.~\ref{fig:viscosity}, demonstrates that at high temperature, the viscosity coefficient very well obeys the   $\theta_0^{3/2}$ scaling, in agreement with predictions~\cite{BruunViscousNormalDamping}. We note that Eq.~\ref{eq:scaling} predicts that
$\alpha_0$ scales nearly as $E^3$, because $\theta_0\propto T_0/n_0^{2/3}\propto E^2$.
This explains the factor of $\simeq 10$ increase in the viscosity coefficients as the initial energy is increased from $E=2.3\,E_F$ to $E=4.6\,E_F$.

A precise comparison between the viscosity data and theory requires calculation of the trap-average $\bar{\alpha}$ from the local shear viscosity, where the relation is tightly constrained by the observed $T^{3/2}$ scaling. Our simple approximation $\bar{\alpha}\simeq \alpha_0$, yields  $\alpha_{3/2}=3.4(0.03)$, where $0.03$ is the statistical error from the fit. A better estimate based on a relaxation model~\cite{Schaefer:2009px} shows that  $\bar{\alpha}= 1.3\,\alpha_0$ at high $T$, yielding $\alpha_{3/2}=2.6$. At sufficiently high temperature, the mean free path becomes longer than the interparticle spacing, since the unitary collision cross section decreases with increasing energy. In this limit, a two-body Boltzmann equation description of the viscosity is valid.  For a Fermi gas in a 50-50 mixture of two spin states, a variational calculation~\cite{BruunViscousNormalDamping} yields $\alpha_{3/2}=45\,\pi^{3/2}/(64\sqrt{2})=2.77$, in reasonable agreement with the fitted values.

\begin{figure}
\begin{center}\
\includegraphics[width=120mm]{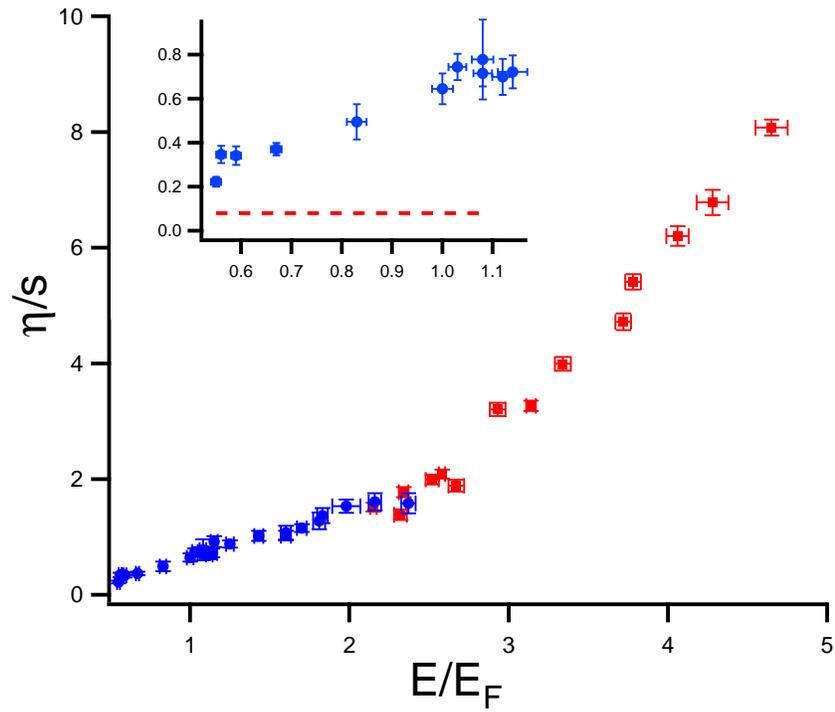}
\end{center}
\caption{Estimated ratio of the shear viscosity to the entropy density. Blue circles: Breathing mode measurements; Red squares: Anisotropic expansion measurements; Inset: Red dashed line denotes the string theory limit.  Bars denote statistical error arising from the uncertainty in $E$, $\bar{\alpha}$, and $S$~\cite{SupportOnline}.\label{fig:etaovers}}
\end{figure}

Finally, Fig.~\ref{fig:etaovers} shows an estimate of the ratio of $\eta/s=\alpha\hbar n/s=(\hbar/k_B)\alpha/(s/nk_B)\simeq(\hbar/k_B)\bar{\alpha}/S$, where $S$ is the average entropy per particle of the trapped gas in units of $k_B$. We obtain $S$ in the low temperature regime from Ref.~\cite{ThermoLuo}, which joins smoothly to the second virial coefficient approximation for $S$ in the high temperature regime~\cite{SupportOnline}. The inset shows the low temperature behavior, which is about five times the string theory limit (red dashed line) near the critical energy~\cite{ThermoLuo} $E_c/E_F = 0.7-0.8$~\cite{SupportOnline}.  We note also that the apparent decrease of the $\eta/s$ ratio as the energy approaches the ground state~\cite{ThermoLuo} $0.48\,E_F$ does not require that the local ratio  $\rightarrow 0$ as $T\rightarrow 0$, since contributions from the cloud edges significantly increase $S$  compared to the local $s$ at the center.

%\bibliography{119521BibTexData}

\begin{thebibliography}{10}
\bibitem{p-wave}
The experiments are performed far from p-wave Feshbach resonances. The relevant
  threshold energy for p-wave scattering is then comparable to the barrier
  height. Using the known $C_6$ coefficients, the barrier height for $^{40}$K
  is $280\,\mu$K, while for $^6$Li, it is 8 mK. Hence, for temperatures in the
  $\mu$K range, as used in the experiments, p-wave scattering is negligible and
  s-wave scattering dominates.
\bibitem{DampNote}
We give the damping rate $1/\tau$ for a cylindrically symmetric cigar-shaped
  trap. For $\delta\equiv(\omega_x-\omega_y)/\sqrt{\omega_x\omega_y}<<1$, with
  $\omega_x,\omega_y$ the transverse trap frequencies, $1/\tau$ contains an
  additional factor $1-\delta$.
\end{thebibliography}

\begin{scilastnote}
\item This research is supported by the Physics Divisions of the
 National Science Foundation, the Army Research Office, the Air Force Office Office of Sponsored Research, and the
Division of Materials Science and Engineering,  the
Office of Basic Energy Sciences, Office of Science, U.S.
Department of Energy. T. S. and J. E. T. are affiliated with the ExtreMe Matter Institute (EMMI).
\end{scilastnote}

\noindent \textbf{Supporting Online Material}\\
www.sciencemag.org\\
Materials and Methods\\
Fig. S1, S2\\
Supporting References and Notes

\newpage

\end{document}